\documentclass{aa}

\usepackage{graphicx}
\usepackage{txfonts}
\usepackage[rflt]{floatflt}
\usepackage{natbib}        
\bibpunct{(}{)}{;}{a}{}{,} 

\newcommand\dtk{IONIC-2TK}
\newcommand\dth{IONIC-2TH}

\begin{document}
\title{Integrated optics for astronomical interferometry}
\subtitle{VI. Coupling the light of the VLTI in K band\thanks{Based on observations collected at the European Southern Observatory, Paranal, Chile (public commissioning data)}}
\author{
  J.-B.~LeBouquin\inst{1}
  \and P.~Labeye\inst{2}
  \and F.~Malbet\inst{1}
  \and L.~Jocou\inst{1}
  \and F.~Zabihian\inst{1}
  \and K.~Rousselet-Perraut\inst{1}
  \and J.-P.~Berger\inst{1}
  \and A.~Delboulb\'e\inst{1}
  \and P.~Kern\inst{1}
  \and A.~Glindemann\inst{3}
  \and M.~Sch\"oeller\inst{4}}
\institute{
  LAOG - Laboratoire d'Astrophysique UMR UJF-CNRS 5571, Observatoire de
  Grenoble, Universit\'e Joseph Fourier, BP 53, 38041 Grenoble, France
  \and LETI - Laboratoire d'Electronique de Technologie et d'Information,
  CEA 17 rue des Martyrs, 38054 Grenoble, France
  \and European Southern Observatory, Karl-Schwarzschild-Str. 2,
  85748 Garching, Germany
  \and European Southern Observatory, Casilla 19001, Santiago 19, Chile
}
\offprints{J.B.~LeBouquin\\
  \email{jean-baptiste.lebouquin@obs.ujf-grenoble.fr}}
\date{Accepted 20 December 2005}
\abstract {
} {
  Our objective is to prove that integrated optics (IO) is not only a good concept for astronomical interferometry but also a working technique with high performance.
} {
  We used the commissioning data obtained with the dedicated K-band integrated optics two-telescope beam combiner which now replaces the fiber coupler MONA in the VLTI/VINCI instrument. We characterize the behaviour of this IO device and compare its properties to other single mode beam combiner like the previously used MONA fiber coupler.
} {
  The IO combiner provides a high optical throughput, a contrast of $89$\% with a night-to-night stability of a few percent. Even if a dispersive phase is present, we show that it does not bias the measured Fourier visibility estimate. An upper limit of $5\times10^{-3}$ for the cross-talk between linear polarization states has been measured. We take advantage of the intrinsic contrast stability to test a new astronomical prodecure for calibrating diameters of simple stars by simultaneously fitting the instrumental contrast and the apparent stellar diameters. This method reaches an accuracy with diameter errors of the order of previous ones but without the need of an already known calibrator.
} {
  These results are an important step of integrated optics, since they prove its maturity in an astronomical band where the technology has been specially developed for astronomical conveniences. It paves the road to incoming imaging interferometer projects.
}
\maketitle


\section{Introduction}

Installed at the heart of the Very Large Telescope Interferometer \citep[VLTI,][]{Glindemann-2003}, the VINCI instrument combines coherently the light coming from two telescopes in the infrared K band. Among the most impressive astrophysical results is the measurements of diameters of very low mass stars \citep{segransan-2003}, the oblateness of the fast rotating star Achernar \citep{Domiciano-2003}, the calibration of the brightness-color relation of Cepheids \citep{Kervella-2004b} and its complementarity with asteroseismology to constrain the stellar structure \citep{Pijpers-2003}. All of them benefit from high accuracy interferometric measurements, achieved in the near-infrared by modal filtering of the corrugated wavefront and real time monitoring of the stellar flux injection \citep{Foresto-1997,Tatulli-2004}.

\citet[paper I:][]{Malbet-1999} have suggested to combine beams with planar integrated optics (IO) components to take benefit of strong spatial filtering, stability and compactness. Afterwards, this technique has been successfully tested in laboratory \citep[papers II and III:][]{Haguenauer-2000b,Berger-1999} and on the sky \citep[paper IV:][]{Berger-2001}.

In the framework of the IONIC activities, a collaboration was initiated in 2003 between LAOG and the ESO interferometry group to implement an integrated optics beam combiner operating in the K band on the VLTI, sharing the VINCI optical interface. This proposal followed the study of the IO techniques towards the K band \citep[paper V:][]{Laurent-2002} and a previous collaboration between the two institutes concerning an IO H band combiner \citep{LeBouquin-2004}.

\begin{figure*} 
  \centering
  \includegraphics[width=0.9\textwidth]{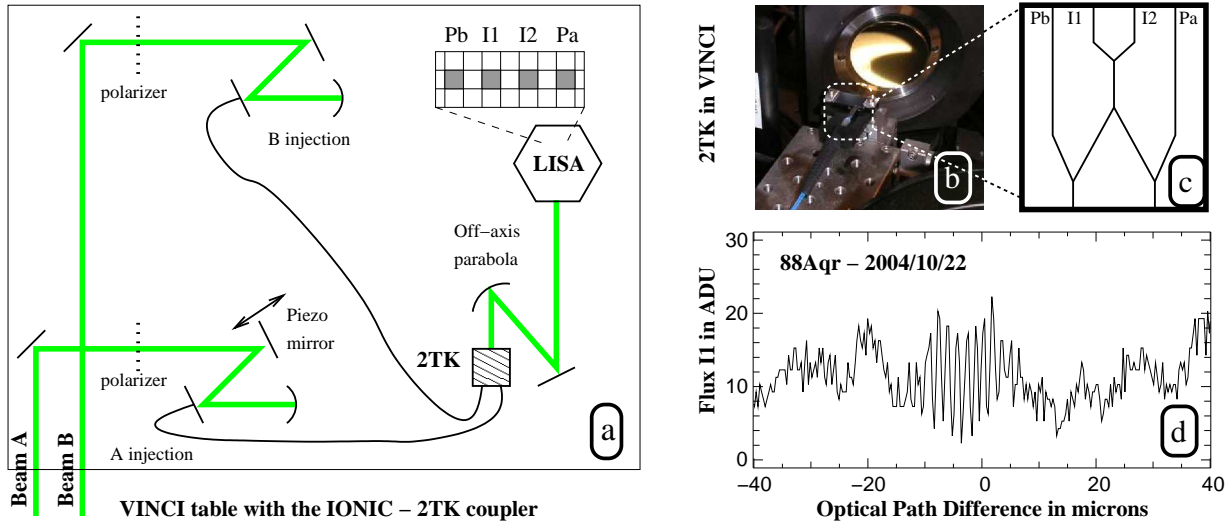}
   \caption{ {\bf a:} Layout of the VINCI instrument where the MONA fiber coupler is replaced by the \dtk{} Integrated Optics coupler. The LISA detector array has been expanded to show the four pixels: Pa and Pb monitor the photometry and the fringes are recorded on I1 and I2. {\bf b:} Picture of the \dtk{} coupler in front of the VINCI off-axis parabola. {\bf c:} Sketch of the component with the two inputs (bottom) and the four outputs (top). {\bf d:} Raw interferometric output I1 obtained in the K band on 88~Aqr with the siderostats. Note the large fringe contrast even though the photometry is not calibrated.}
\label{fig:implantation_2TK}
\end{figure*} 

In this paper, we report the validation of the new VINCI setup, equipped with a two-telescope IO coupler for the K band. In Sect.~\ref{sec:commissioning}, we summarize the instrumental context, the observations and the data reduction techniques. Results are presented and discussed in Sect.~\ref{sec:results}. The intrinsic instrumental stability allows us to validate a new interferometric calibration technique, based on the simultaneous fit of the instrumental contrast and stellar apparent diameter as detailed in Sect.~\ref{sec:new_contrast}. Finally, perspectives of Integrated Optics in the framework of optical interferometry are discussed in Sect.~\ref{sec:conclusions}.


\section{Observational context}
\label{sec:commissioning}

\subsection{The VINCI + \dtk{} instrument}

VINCI combines coherently the light coming from two telescopes of the VLTI array as detailed in \citet{Kervella-2000}. Figure~\ref{fig:implantation_2TK} shows the sketch of VINCI: the two beams enter the instrument from the bottom of the figure. The light is injected into the fibers and then combined in a single mode coupler. The original device is a fibered coupler called MONA based on the same principle as the FLUOR instrument \citep{Foresto-1998}. In 2003, an integrated optics combiner has been manufactured to allow operation in the H band \citep[\dth{},][]{LeBouquin-2004}. 
The optical path difference (OPD) between the two beams is modulated to sweep through the interference fringes, that appear as temporal modulation. During the observations, a simple fringe packet centroiding algorithm is applied, removing the instrumental OPD drift.

The new integrated optics coupler (\dtk{}) has been manufactured by the LETI\footnote{CEA-LETI: \texttt{http://www-leti.cea.fr}} with a silica on silicon technology. It is connected at the input by polarization-maintaining single mode fibers. Both the fibers and the component waveguide geometry are optimized for the K band, which is not the standard wavelength in telecommunications. On the VINCI table, the chip replaces the bundle of fibers coming from the MONA box. Injection fibers of \dtk{} are fed by the off-axis parabolas and the component output is directly imaged on the detector (see Fig.~\ref{fig:implantation_2TK}). The camera software has been modified to allow aligned pixel reading. Perfect superposition of the four spots into four pixels is impossible, due to an unsuitable magnification of the output optics. Since only the central pixel is read for each spot, a fraction of light is lost. Yet it has no drastic influence on the VINCI operations.

\subsection{Observations and data reduction}

\begin{table}
  \centering
  \caption{List of VINCI with \dtk{} observations with the 35~cm siderostats. The diameters are extracted from the CHARM2 catalog \citep{Richichi-2005}.}
  \begin{tabular}[c]{c|cccc}\hline\hline
    Date of & Target & K     & Nb. of & UD \diameter \\
    obs.    & name   & [mag] & obs.   & [mas] \\\hline
    2004-08-19 & Internal Light        &  & 20 & with polarizers \\ 
    2004-08-22 & $\lambda$~Sgr         &  0.332 & 2 & 4.36 $\pm$ 0.21\\    
    -          & $\theta$~Cen          & -0.274 & 1 & 5.32 $\pm$ 0.06\\
    -          & $\alpha$~Eri          &  0.880 & 2 & 1.91 $\pm$ 0.15\\
    -          & \object{$\alpha$~PsA} &  0.945 & 2 & 2.19 $\pm$ 0.02\\
    -          & $\chi$~Aqr            & -0.365 & 2 & 6.70 $\pm$ 0.15\\
    -          & $\epsilon$~Sco        & -0.392 & 6 & 5.83 $\pm$ 0.06\\\hline
    2004-10-08 & \object{88~Aqr}       &  0.986 & 23 & 3.24 $\pm$ 0.20\\
    -          & \object{$\alpha$~PsA} &  & 23 \\
    2004-10-09 & \object{88~Aqr}       &  & 23 \\
    -          & \object{$\alpha$~PsA} &  & 22 \\
    2004-10-10 & \object{88~Aqr}       &  & 21 \\
    -          & \object{$\alpha$~PsA} &  & 19 \\
    2004-10-11 & \object{88~Aqr}       &  & 8 \\
    -          & \object{$\alpha$~PsA} &  & 6 \\\hline
  \end{tabular}
  \label{tab:resume_obs_2TK}
\end{table}

Table~\ref{tab:resume_obs_2TK} summarizes the K band observations used to qualify VINCI with the \dtk{} coupler. During the commissioning (August 2004), a strong flux unbalance appeared between the two siderostats. Therefore we observed only bright and well known stars, with many different apparent diameters. In October 2004 the VLTI team observed 88~Aqr and \object{$\alpha$~PsA} consecutively for few days. Results presented in this paper also use previous observations done with MONA and \dth{} couplers in the K and H band respectively. The assumed uniform disk diameters are extracted from the CHARM2 catalog \citep{Richichi-2005} except for \object{$\alpha$~PsA} which has been measured by \citet{DiFolco-2004} with a good accuracy.

We follow the data reduction procedure detailed in \citet{LeBouquin-2004}, based on the Fourier technique introduced by \citet{Foresto-1997}. It gives access to the photometric extraction coefficients $\kappa$, the fringe contrast in the direct space $\overline{\mu}$ and the fringe energy in the Fourier space $\overline{\mu^2}$.


\section{Results}
\label{sec:results}

As far as an astronomer is concerned, the instrumental quantities of interest when dealing with a two-telescope interferometer are the optical throughput, the instrumental contrast, its stability and its chromatic and polarized response.

\subsection{Optical throughput and limiting magnitude}
\label{sec:throughput}

Including only coupling losses and component transmission, the throughput of the \dtk{} combiner measured at the LAOG test bench reaches 70\%. We also compare the recorded flux on each star with previous measurements obtained with the MONA coupler. The transmission of the VINCI + \dtk{} setup is at least equal to the VINCI + MONA one. The same un-balance between beam A and beam B is observed, which points out to a difference between the two arms before the focal instrument. The limiting magnitude of VINCI with the \dtk{} combiner is thus equal to the previous VINCI + MONA in the K band. With this last setup, \citet{Wittkowski-2004b} observed NGC~1068 ($K=9.2$) with the 8 meters telescopes equipped with adaptive optics. Since the signal-to-noise ratio were good although a short exposure time, we can expect a VINCI limiting magnitude better than $K=10$. The targets observed in this paper are far from this magnitude especially because we used the 35 cm siderostats and because of the flux un-balance, by now already fixed.

\subsection{Intrinsic contrast and stability}
\label{sec:contrast}

Testing the stability of the photometric extraction coefficients $\kappa$ is not straightforward, since the obtained value takes into account the mixed contributions of the component, the opto-mechanic and the detector. Nevertheless, the measured coefficients are not correlated with the target, showing no chromatic dependence. The data dispersion during a typical night is within the error bars (about 1\%) and the differences from night to night are explained by optical re-alignments \citep{LeBouquin-2004}.
\begin{figure*}
  \centering
  \begin{minipage}{0.49\hsize} 
    \centering
      \includegraphics[angle=0,width=\textwidth]{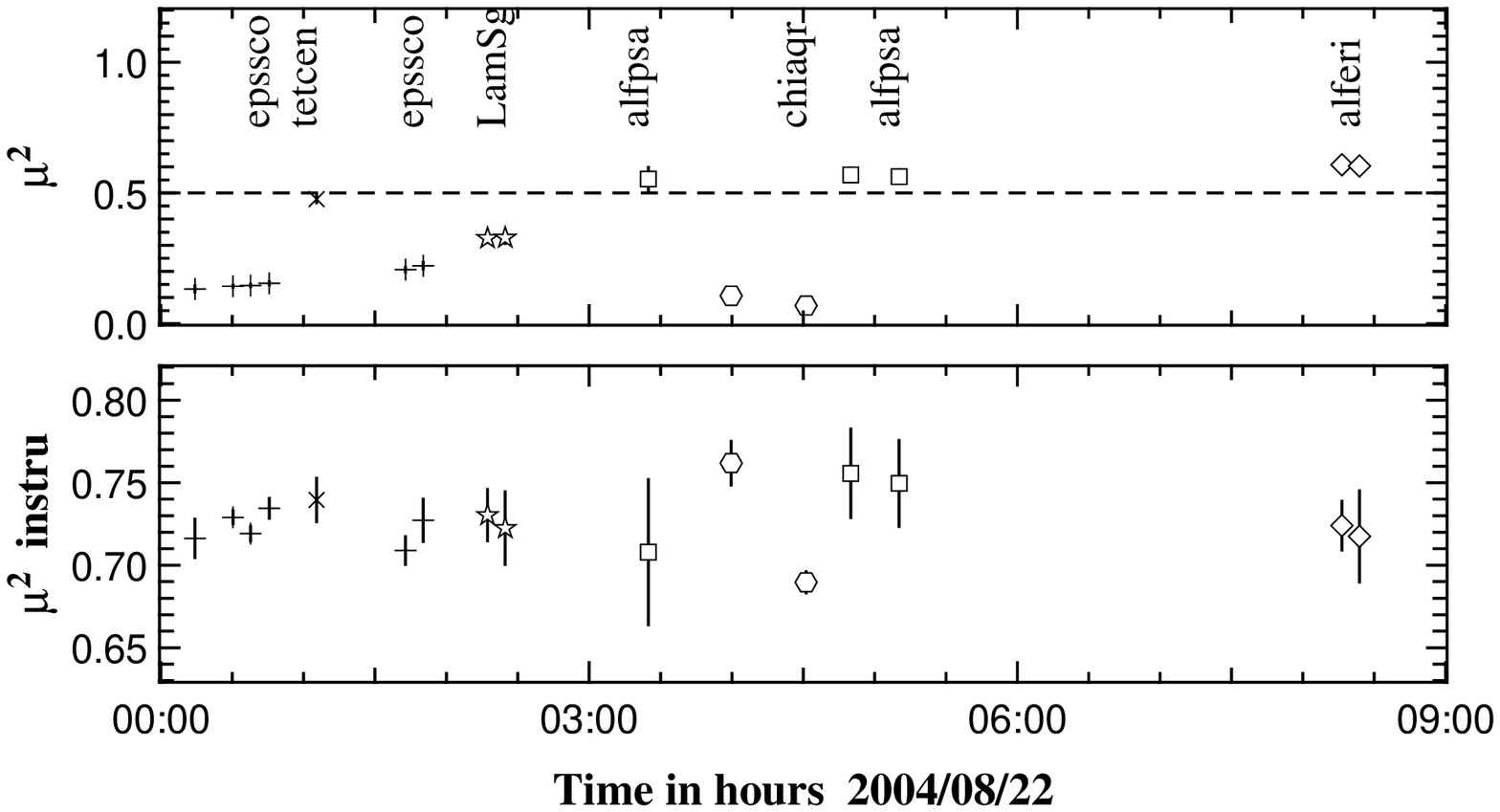}
  \end{minipage}
  \begin{minipage}{0.49\hsize} 
    \centering
    \includegraphics[angle=0,width=\textwidth]{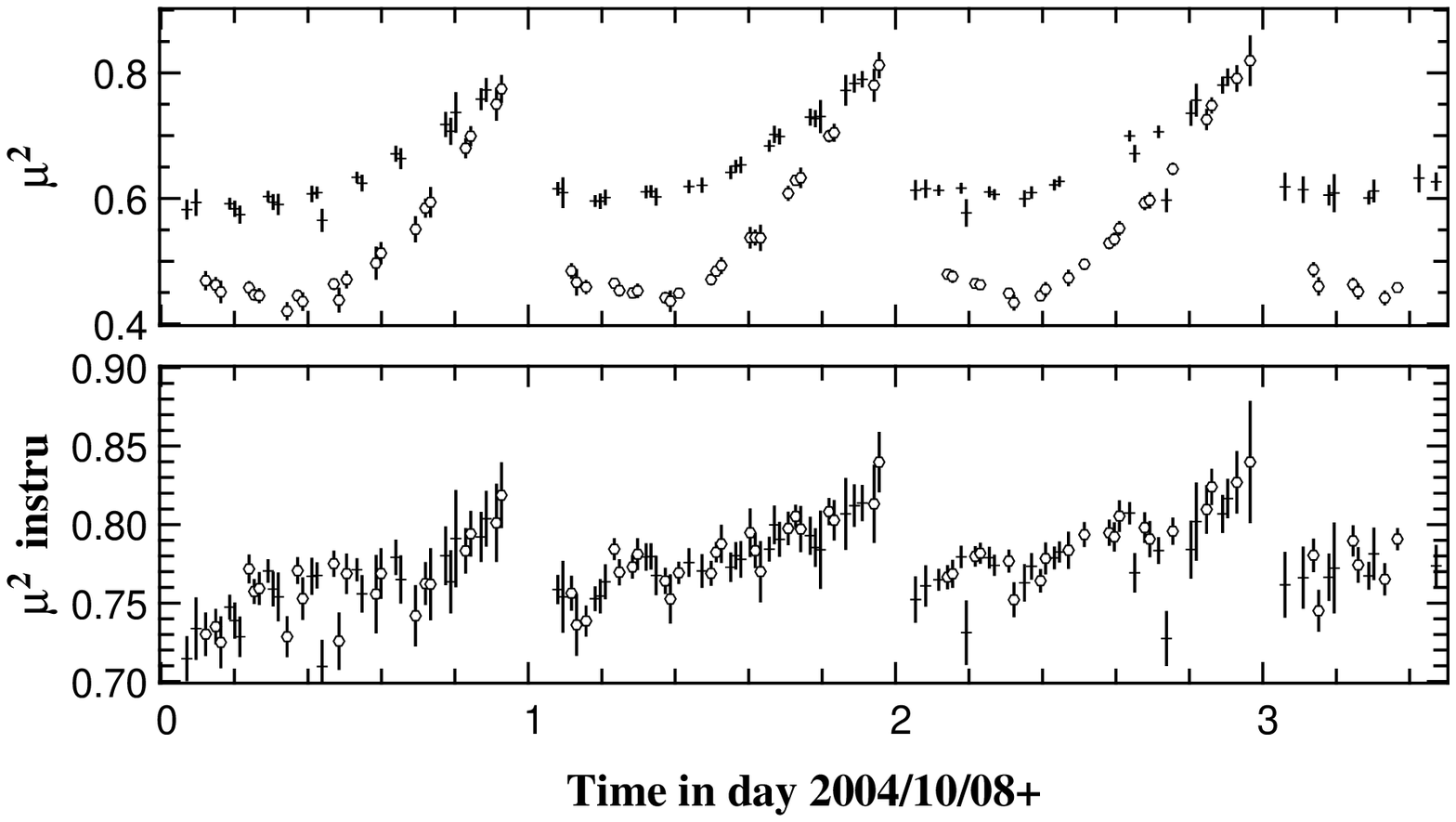}
  \end{minipage}
  \caption{\emph{Fourier estimator} of the square visibilities ($\overline{\mu^2}$, top) and corresponding instrumental square visibilities (bottom) observed with VINCI + \dtk{} and the siderostats. The data of October 2004 on 88~Aqr (circles) and $\alpha$~PsA (crosses) for four consecutive nights show the effect of the earth rotation synthesis. We assume an effective wavelength of $\lambda_0=2.178\mu$m and the uniform disk angular diameters of Table~\ref{tab:resume_obs_2TK}. When not seen, errors bars are within the symbol sizes.}
  \label{fig:2TK_contrasteSky}
\end{figure*}

For each observation, the square instrumental contrast is estimated by dividing the measured square visibility (Fourier estimator) by the expected one on stars with known diameter. We take an effective wavelength of $\lambda_0=2.178\mu$m \citep{Kervella-2004} and the uniform disk angular diameters of Table~\ref{tab:resume_obs_2TK}. The dispersion is compatible with the errors bars (Fig.~\ref{fig:2TK_contrasteSky}, left). The instrument is stable from night to night (Fig.~\ref{fig:2TK_contrasteSky}, right). The variations can be explained by different night qualities and are strongly correlated to the atmospheric coherence time recorded at Paranal. The instrumental contrast during the commissioning week of August 2004 was $84$\% (corresponding to $\mu^2_{instru}=0.7$) and reaches $89$\% (corresponding to $\mu^2_{instru}=0.8$) in the October 2004 data, to be compared with the $95$\% measured in the LAOG laboratory. Between these two data sets, the total flux and the beam balance have been strongly improved, pointing that the instrumental contrast evolution was most probably due to the VLTI than to the combiner.

\subsection{Chromatic transmission and phase}
\label{sec:chromatic}

\begin{figure} 
  \centering
  \leavevmode
  \includegraphics[angle=0,width=0.49\textwidth]{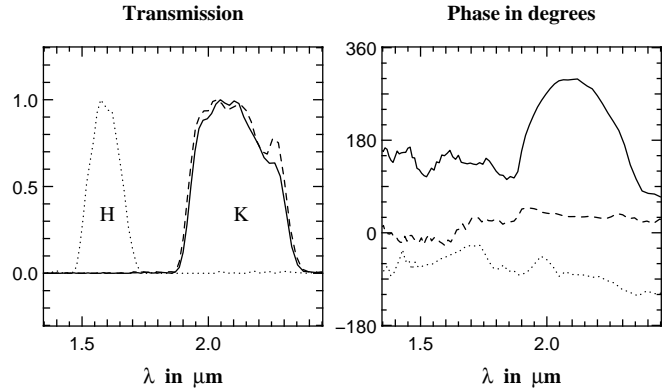}
  \caption{Chromatic transmission (left) and phase (right) of VINCI analyzed in Fourier Transform Spectrometer mode. The couplers used are \dth{} (dot, 2002/07/23), MONA (dash, 2004/07/19) and \dtk{} (solid, 2004/08/19). The transmission is normalized by its maximum in the considered band.}
  \label{fig:FTS_2TK}
\end{figure}
With the internal light, VINCI becomes a simple Michelson interferometer and can be used as a Fourier Transform Spectrometer to explore its own chromatic response. Spectra for different instrumental setups are displayed in Figure~\ref{fig:FTS_2TK}. These curves depend on the combiner and fiber transmissions, but also on the filter used and the spectral behavior of the internal light. Therefore, we focus on differences between the setups.

\begin{figure}
  \centering
  \includegraphics[angle=0,width=0.5\textwidth]{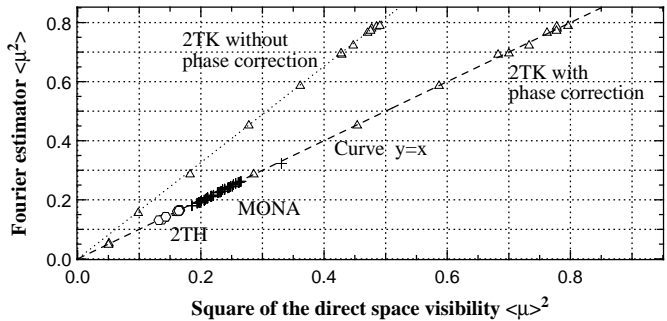}
  \caption{\emph{Fourier estimator} $\overline{\mu^2}$ versus \emph{direct space estimator} $\overline{\mu}^2$ of the square visibility for different VINCI setups fed with the internal light. With the \dtk{} coupler, the direct space estimation is performed with and without dispersion correction (see Sect.~\ref{sec:chromatic}). Error bars are within the symbol size. The dash curve is $y=x$.}
  \label{fig:mumu2_2TK}
\end{figure}

The transmitted spectral shape is similar for MONA and \dtk{} coupler (left panel). The phase of MONA and \dth{} are really constant over the transmitted band while it is no longer true for the \dtk{} (right panel). Its parabolic profile points out to dispersion. The optical path difference between the two beams inside the IO combiner should be very small, thanks to a manufacturing accuracy better than a fraction of $\lambda$. Besides, the large visibility obtained with the LAOG bench before the fiber connecting ($>95$\%) allows us to definitely exonerate the IO combiner. Thus, the dispersion is most probably due to the input fibers. Using the expression given by \citet{Foresto-1995} and a standard dispersive coefficient of $D=50$~ps/nm/km for the silica, we find an optical length difference of $\Delta L = 9$~mm. This particularly large value cannot be explained only by a geometrical deformation of the fibers. Probably there is also a difference of the silica optical index due to mechanical constraints on the fibers.

Figure~\ref{fig:mumu2_2TK} shows that MONA and \dth{} match a relation close to $\overline{\mu^2}=\overline{\mu}^2$. It is not yet true for \dtk{}. We re-compute the direct space visibilities with a \emph{dispersion corrected} estimator. The phase under the Fourier peak at the fringe frequencies is removed by a second order polynomial fit, before taking the maximum of the packet envelope in the direct space. The Fourier square visibility $\overline{\mu^2}$ is not modified by this operation. When correcting the dispersion, the \dtk{} combiner catches up the relation $y=x$ (Figure~\ref{fig:mumu2_2TK}).

The contrast is thus modified \emph{only} by a dispersive phase, which does not affect the classical Fourier estimators. The standard data reduction software of VINCI is therefore not biased. Nevertheless, attention has to be paid on the wavelet algorithm \citep{Kervella-2004}, because some selection criteria use the spectro-temporel fringe shape, which is affected by the dispersion. The spread of the fringe packet in the OPD space may reduce the efficiency of the VINCI fringe tracking method. In practice, this is not true because the direct space contrast remains better with \dtk{} than with MONA. We checked that the spread of the fringe packet is not large enough to bias the Fourier estimation because of the finite OPD modulation.

\subsection{Polarization}
\label{sec:polarisation}

When rotating polarizers in Beam A, the integrated flux on $5\times5$ pixels around the Pa output spot remains constant, while the flux on the central pixel changes by a factor 3. This behavior is repeatable and correlated to the input polarization angle. A small difference of a fraction of pixel of spot position between the two polarization states could explain the variations. This effect is never seen on the other outputs or with the polarizer in Beam B. The classical photometric calibration removes its influence on the data.

Placing the two polarizers in the two input beams allows to explore the polarization properties of VINCI. The polarization angle on Beam B was always set to 225$^\circ$ (direction of source polarization). By rotating the polarizer in Beam A, the square visibility decreases to $\overline{\mu^2}=0.005$. It corresponds to an equivalent polarization cross-talk that could be explained by the non-perfect extinction of the polarizers for wavelengths smaller than $2\mu$m and by neutral axis alignment. $5\times10^{-3}$ is thus an upper limit accuracy of the VINCI + \dtk{} polarization cross-talk.

\subsection{Behavior of the Integrated Optics coupler}
\label{sec:io_component}

A faint fringe packet has been observed in the two photometric outputs Pa and Pb of the 2TK coupler when modulating the OPD and when flux is injected in the two inputs. The packet is at the same position as the scientific fringes recorded in the interferometric outputs I1 and I2. Its contrast of approximatively 1\% points out to a flux cross-talk inside the component at the $10^{-4}$ level. These fringes have no incidence on the VINCI results since they are removed by the Wiener filtering in the standard data reduction procedure described by \citet{Kervella-2004}.


\section{Self determination of stellar diameters}
\label{sec:new_contrast}

\begin{figure} 
  \centering
  \includegraphics[width=.47\textwidth]{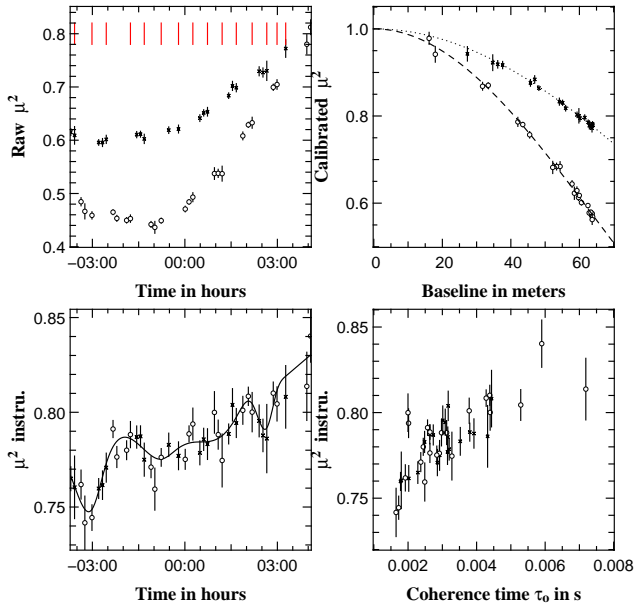}
  \caption{Simultaneous fit of the system contrast (solid line) and the stellar diameters of 88~Aqr (circles, dash lines) and $\alpha$~PsA (crosses, dot line). {\bf Top-Left:} Raw square visibilities versus time. The degrees of freedom of the spline function are represented by the vertical solid lines. {\bf Bottom-Left:} Recovered instrumental square contrast and results of the fit versus time. {\bf Top-Right:} Recovered calibrated square visibilities and results of the fit versus the projected baseline. {\bf Bottom-Right:} Recovered instrumental square contrast versus the coherence time $\tau_0$.}
  \label{fig:Contrast_Base_2TK}
\end{figure}
With the following assumptions, it becomes possible to recover simultaneously the system contrast \emph{and} the stellar diameters without the need of external calibrators~:
\begin{itemize}
\item Different objects should be observed and they should be well modeled with uniform disks.
\item The visibilities should sample a large range of spatial frequencies.
\item The instrumental contrast should remain identical, or at least strongly correlated, for visibilities taken successively on the different objects. Therefore the instrumental contrast can be modeled with a few number of degrees of freedom.
\end{itemize}

Taking advantage of the high internal stability, we propose to test this technique on data obtained with the new VINCI setup during the 2004 October 9 night. We used two models of the atmospheric contrast~: \#1 a linear relation with time (two degrees of freedom) ; \#2 a spline function with 14 degrees of freedom, about one per object switch. Like this the atmospheric contrast is constrained by the observation of the two stars.

\begin{figure} 
  \centering
  \includegraphics[width=.47\textwidth]{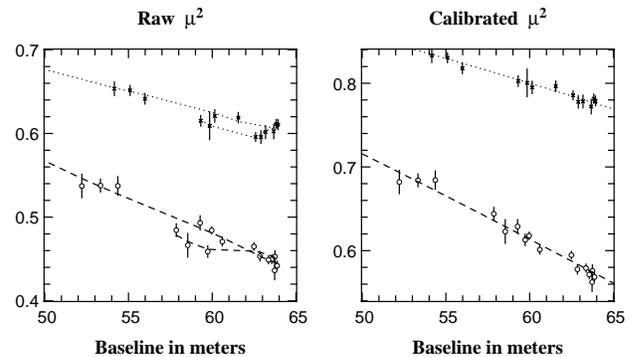}
  \caption{Raw square visibilities (Left) and calibrated square visibilities (Right) versus the projected baseline in meter for 88~Aqr (circles, dash lines) and $\alpha$~PsA (crosses, dot line). The lines corresponds to the recovered uniform disk diameters (right) and the same divided by the recovered ``real-time'' square instrumental contrast (left, this line is function of baseline length and Hour Angle).}
  \label{fig:Contrast_Base_Fit}
\end{figure}

We use the Fourier estimator of the visibilities, we assume an effective wavelength of $\lambda_0=2.178\mu$m and we start with unrealistic diameters of $0.1$mas. The best fit parameters are summarized in Table~\ref{tab:fit_contrast} and results of model \#2 are displayed in Figure~\ref{fig:Contrast_Base_2TK}. Error bars contain both the statistic and systematic errors.
\begin{table} 
  \centering
  \caption{Best simultaneous fit of the uniform disk diameters. The reduced $\chi^2$ is defined as the square distance between measure and best fit divided by the difference between number of degrees of freedom and number of measures.}
  \begin{tabular}[c]{l|cc|c} \hline\hline
     Model    & \object{$\alpha$~PsA}
               & \object{88~Aqr} & $\chi^2$ \\ \hline
                        Previous  &$2.19\pm0.02$& $3.24\pm0.20$ \\
     \#1: Linear $\mu^2_{instru}$ &$2.25\pm0.06$& $3.26\pm0.05$& $1.1$ \\
     \#2: Spline $\mu^2_{instru}$ &$2.23\pm0.07$& $3.26\pm0.06$& $0.51$\\\hline
  \end{tabular}
  \label{tab:fit_contrast}
\end{table}
The reduced $\chi^2$ clearly decreases between model \#1 and \#2, showing that the instrumental contrast is better recovered with the spline function (Fig.~\ref{fig:Contrast_Base_2TK}, bottom-left). These variations can be explained by the evolution of the coherence time during the night (Fig.~\ref{fig:Contrast_Base_2TK}, bottom-right). When taking into account these variations, the visibilities of both objects versus the baseline catch up the slope of a uniform disk (Fig.~\ref{fig:Contrast_Base_Fit}). The best accuracy is achieved with model \#1, but results are probably less biased with model \#2. Complete theoretical study of the method is beyond the scope of this paper, and we focus on the astrophysical results. In agreement with previous measurement of \citet{DiFolco-2004} (our works are completely consistent in terms of wavelength, baselines and hour angles), we found a stellar diameter of \object{$\alpha$~PsA} about 6\% larger than the one measured by \citet{Davis-2005}. For \object{88~Aqr}, we found a uniform disk diameter of $3.26\pm0.05$mas, in agreement with previous photometric estimations and with the $3.22\pm0.04$mas that we obtained with the classical calibration method (assuming a uniform disk diameter of $2.19\pm0.02$mas for \object{$\alpha$~PsA} as calibrator).

The main interest of this technique is to provide a new estimation of the recovered quantities~: instrumental contrast and stellar diameter does not depend on previously measured diameter. It can be applied only to close stars, partially resolved and with largest super-synthesis effect. A first application could be to check the diameter consistency of important interferometric calibrators. Because this method requires a strong intrinsic stability, the new VINCI + \dtk{} setup is a well suited instrument, as illustrated in this paper.


\section{Conclusions and perspectives}
\label{sec:conclusions}

We have characterized the new setup of the VINCI instrument, equipped with an Integrated Optics component for the K band. First, we show that VINCI keeps the same limiting magnitude. Coupled with a good stability and a high instrumental contrast, it makes VINCI an interesting instrument to perform observations on faint objects. Then, we have explored its chromatic response by Fourier Transform Spectrometry. A dispersive phase due to the fibers is present but does not bias the measured square visibility by the classical Fourier estimation. Nevertheless, careful attention has to be paid to the selection criteria of the Wavelet algorithm developed by \citet{Kervella-2004} because the spectro-temporal fringe shape is slightly spread over more frequencies and times. Finally, we find an upper limit of $5\times10^{-3}$ for the cross-talk between linear polarization states.

The intrinsic stability of the whole instrumental chain VLTI + VINCI + \dtk{} allows us to try a new calibration technique, based on the simultaneous fit of the atmospheric contrast and stellar apparent diameter. We validate it with two well known stars. The recovered fit parameters are all consistent. We emphasize that this technique thus proves the interferometric quality and stability of VINCI.

These results are an important step for the development of Integrated Optics for astronomical interferometry. Previous works were all performed to validate the technologies or to test the performances in shared risk programs. For the first time, a component has been designed, manufactured and commissioned to answer an astrophysical institute request. Because of the lack of telecommunication or metrology applications at $2\mu$m, the K band component used has been especially developed for astronomical convenience, proving the maturity of the technique.

In the future, the goal is to combine the entire VLTI array and to disperse the light in order to have spectro-imaging capabilities. Integrated optics is certainly a promising solution \citep{Kern-2003}. The compactness of the planar optical component allows one to combine many beams in the same chip, which drastically reduces the instability and the required alignments. The observational strategies (number of baselines, wavelength, combination scheme...) can be adapted to the object thanks to the ``plug and play'' ability of IO combiners. Output beams of the planar component can act as the input slit of a spectrograph, avoiding complex anamorphic optics.

In this context, we develop a IO chip which combines four beams with a very photon efficient concept: it allows to measure the six complex coherencies with only 24 pixels and without external OPD modulation \citep{LeBouquin-2004b}. This component is already under tests at the LAOG optical bench and could be a key part for incoming imaging interferometer projects such as VITRUV \citep{Malbet-2004}.
 

\begin{acknowledgements}
This work was partially funded by the French spatial agency CNES and by the CNRS/INSU. The authors want to thank the ESO support for the observations, especially Fredrik Rantakyro, Martin Vannier and Bertrand Bauvir. This work is based on observations made with the European Southern Observatory telescopes obtained from the ESO/ST-ECF Science Archive Facility. This research has also made use of  the SIMBAD database at CDS, Strasbourg (France) and the Smithsonian/NASA Astrophysics Data System (ADS). All the calculations and graphics were performed with the freeware \texttt{Yorick}\footnote{\texttt{ftp://ftp-icf.llnl.gov/pub/Yorick/doc/index.html}}.
\end{acknowledgements}



\end{document}